\documentclass[letterpaper, 10 pt, conference]{ieeeconf}  

\IEEEoverridecommandlockouts                              
                                                          
\usepackage{cite,color}
\usepackage{amsmath,amssymb,amsfonts}
\usepackage{comment}
\usepackage{graphicx}

\newtheorem{secthm}{Theorem}[section]
\newtheorem{seccor}[secthm]{Corollary}

\newtheorem{secrem}[secthm]{Remark}

\newcommand{\bR} { {\mathbb R}}

\newcommand{\cC} { {\mathcal C}}

\newcommand{\cO} { {\mathcal O}}

\def\red{\hfill $\lhd$}

\title{\LARGE \bf
On Differential Controllability and Observability Functions*
}

\author{Yu Kawano$^{1}$, Bart Besselink$^{2}$, and Jacquelien M.A. Scherpen$^{3}$
\thanks{*The work of Y. Kawano was partially supported by JST FOREST Program Grant Number JPMJFR222E and JSPS KAKENHI Grant Number JP24K00910.}
\thanks{$^{1}$Y. Kawano is with the Graduate School of Engineering, Hiroshima University, Higashi-Hiroshima 739-8527, Japan
        {\tt\small ykawano@ hiroshima-u.ac.jp}}%
\thanks{$^{2}$B. Besselink is with the Bernoulli Institute for Mathematics, Computer Science and Artificial Intelligence, and Jan C. Willems Center for Systems and Control, University of Groningen, Groningen 9700 AK, The Netherlands
        {\tt\small b.besselink@rug.nl}}%
\thanks{$^{3}$J. M. A. Scherpen is with the Faculty of Science and Engineering, and Jan C. Willems Center for Systems and Control, University of Groningen, Groningen 9700 AK, The Netherlands
        {\tt\small j.m.a.scherpen@rug.nl}}%
}

\begin{document}
\allowdisplaybreaks[4]

\maketitle
\thispagestyle{empty}
\pagestyle{empty}

\begin{abstract}
Differential balancing theory for nonlinear model reduction relies on differential controllability and observability functions. In this paper, we further investigate them from two different perspectives. First, we establish novel connections between these differential energy functions and their incremental counterparts by assuming the existence of the corresponding optimal state feedback for each controllability function. Specifically, an upper bound on the incremental controllability/observability function is provided by the corresponding differential energy function. Conversely, an upper bound on the differential controllability function can be estimated from the incremental controllability function. Furthermore, the differential observability function can be constructed from the incremental observability function. Second, we explore the positive definiteness of the differential controllability/observability function in the context of controllability/observability and stability.
\end{abstract}




\section{Introduction}
One of the most well-known approaches to model reduction in linear control systems is balanced truncation \cite{Antoulas:05,Moore:81}. Its core idea is to construct a reduced-order model by retaining the components of a high-order system that contribute most to its input-output behavior. In this sense, balancing theory can be viewed as an extension of the singular value decomposition to control systems. The importance of the state components is evaluated in terms of controllability and observability measured by the controllability and observability Gramians, or equivalently, the controllability and observability functions.

Nonlinear extensions of balancing theory primarily rely on generalizations of the controllability and observability functions. While these nonlinear energy functions are equivalent when restricted to the linear case, they lead to distinct nonlinear balancing approaches, including (standard) nonlinear balancing~\cite{Scherpen:93,FS:05,FS:10}, incremental balancing~\cite{BWS:13,BWS:14}, and differential balancing~\cite{KS:17,KS:21,Kawano:22}. Nonlinear balancing theory is based on (standard) controllability and observability functions defined relative to the distance from an equilibrium point. Incremental balancing theory, inspired by incremental stability analysis~\cite{Angeli:02}, employs energy functions defined by pairs of trajectories of the same system. On the other hand, differential balancing theory, inspired by contraction analysis~\cite{LS:98,FS:14,Bullo:22}, utilizes energy functions defined for the prolonged system, which consists of the system itself and its associated variational system. The main objective of this paper is to establish relations among these energy functions and to provide a deeper understanding of the energy functions, focusing on differential balancing.


\subsubsection*{Literature Review} 
The link between the energy functions for standard nonlinear and incremental balancing is relatively clear. In fact, fixing a trajectory at an equilibrium point in the incremental controllability/observability function recovers the standard controllability/observability function. In contrast, the connection between incremental and differential controllability/observability function remains unexplored, despite the strong relationships between incremental stability and contraction analysis, found in~\cite{FS:14,Bullo:22,AJP:20,KBC:20,WD:22,KB:24} for stability and \cite{VKT:20,KB:24} for passivity. 

On the other hand, the positive definiteness of the controllability/observability Gramian of a linear system is characterized by controllability/observability and stability. These properties are partially extended to nonlinear systems for standard nonlinear balancing~\cite{Scherpen:93,SG:00} and for differential balancing~\cite{KS:17}. Since differential balancing theory intersects with differential geometry, local strong accessibility and observability are utilized for analysis~\cite{NS:15}. However, compared to the linear case, gaps remain in our understanding of the nonlinear case.


\subsubsection*{Contribution} 
The contribution of this paper is twofold. First, we tailor the definitions of differential and incremental energy functions to establish their novel connections, partly drawing on techniques from \cite{KB:24}. Assuming the existence of a suitable optimal feedback, we provide an upper bound on the incremental controllability function based on the differential controllability function, and vice versa. Similarly, we estimate an upper bound on the incremental observability function from the differential one. Notably, we derive the differential observability function directly from the incremental one.

Second, we investigate the positive definiteness of the tailored differential controllability and observability functions. For the differential observability function, we establish a connection among its positive definiteness, zero-state detectability~\cite{Schaft:92}, and the convergence of the corresponding variational system. For analytic systems, zero-state detectability is equivalent to the local observability rank condition~\cite{NS:15} being satisfied. These established relations can also be applied to the analysis of the differential controllability function through the duality between controllability and observability. Specifically, we connect its positive definiteness to the zero-state detectability and convergence of a dual variational system along the trajectory of the closed-loop system under suitable optimal state feedback. In the linear case, we recover the well-known results for the controllability Gramian. However, in the nonlinear case, the zero-state detectability of the dual variational system differs from the local strong accessibility rank condition~\cite{NS:15}, although the zero-state detectability still relates to a Lie bracket condition.

The contributions of this paper are summarized as follows: 
\begin{itemize} 
\item We establish novel connections between the differential and incremental controllability/observability functions; 
\item We provide a deeper understanding of the positive definiteness of the differential observability function by analyzing the zero-state detectability and convergence of the corresponding variational system; 
\item We interpret the positive definiteness of the differential controllability function through the zero-state detectability and convergence of the dual variational system. 
\end{itemize}


\subsubsection*{Organization}
The remainder of this paper is organized as follows.
In Section~\ref{sec:diff}, we recall the definitions of the differential controllability and observabilty functions.
In Section~\ref{sec:energy}, we investigate their relations with the incremental controllability and observabilty functions.
In Section~\ref{sec:pd}, we study the positive definitness of the differential energy functions, which is exemplified in Section~\ref{sec:ex}.
Finally, Section~\ref{sec:con} concludes this paper.


\subsubsection*{Notation} 
The set of real numbers is denoted by $\bR$. The vector space of Lebesgue-measurable, square-integrable functions mapping $[t_0, t_f]$ ($t_0 < t_f$) to $\bR^m$ is denoted by $L_2^m[t_0, t_f]$. For square matrices, $P \succ Q$ (resp. $P \succeq Q$) means that $P - Q$ is symmetric and positive (resp. semi) definite. The vector $2$-norm or the induced matrix $2$-norm is denoted by $|\cdot|$. The weighted vector norm by $P \succ 0$ is denoted by $|x|_P:=\sqrt{x^\top P x}$. The signal $2$-norm is denoted by $\| x \|_{L_2^m[t_0, t_f]} := (\int_{t_0}^{t_f} |x(t)|^2 dt)^{1/2}$.




\section{Preliminaries}\label{sec:diff}
Consider a nonlinear system in the form
\begin{align}\label{eq:sys}
\left\{\begin{array}{r@{}lr@{}l}
\dot x &{} = f(x) + g (x) u, & x(0) = x_0,\\
y &{} = h(x),
\end{array}\right.
\end{align}
where $x \in \bR^n$, $u \in \bR^m$, $y \in \bR^p$, and the functions
$f:\bR^n \to \bR^n$, $g:\bR^n \to \bR^{n\times m}$, and $h:\bR^n \to \bR^p$ are of class~$C^2$.
We let $\varphi (t, x_0, u)$ denote its state solution, at time $t$, starting from $x(0)=x_0$ with input $u$, i.e., $x(t) = \varphi (t, x_0, u)$.

The prolongation~\cite{CSC:05} of~\eqref{eq:sys} is given by
\begin{align}\label{eq:dsys}
\left\{\begin{array}{r@{}l}
\dot x &{} = f(x) + g (x) u \\[1mm]
\dot {\delta x} &{} \displaystyle
= \Biggl( \frac{\partial f(x)}{\partial x} + \sum_{j=1}^m \frac{\partial g_j (x)}{\partial x} u_j \Biggr) \delta x 
+ g (x) \delta u \\[1mm]
y &{} = h(x)\\
\delta y &{} \displaystyle
= \frac{\partial h(x)}{\partial x} \delta x,
\quad (x(0), \delta x(0)) = (x_0, \delta x_0),
\end{array}\right.
\end{align}
where $\delta x \in \bR^n$, $\delta u \in \bR^m$, and $\delta y \in \bR^p$. 

As an energy function of the prolonged system~\eqref{eq:dsys}, the \emph{differential controllability function} \cite[Definition 2.2]{KS:17} is defined by
\begin{align}\label{eq:dcon_func}
E_{d\cC}  (x_0, u, \delta x_0) := 
\inf_{\substack{\delta u \in L_2^m (-\infty, 0]  \\ \delta x(-\infty)=0}} \frac{1}{2} \| \delta u \|_{L_2^m(-\infty, 0]}^2
\end{align}
for $x_0 \in \bR^n$, $u: (-\infty, 0] \to \bR^m$, and $\delta x_0 \in \bR^n$.
Similarly, the \emph{differential observability function} \cite[Definition 2.3]{KS:17} is defined as
\begin{align}\label{eq:dob_func}
E_{d\cO} (x_0, \delta x_0) := \frac{1}{2} \| \delta y \|_{L_2^p[0,\infty)}^2 \; \\
(u(t), \delta u(t)) \equiv (0, 0)
\nonumber
\end{align}
for $x_0 \in \bR^n$ and $\delta x_0 \in \bR^n$.

For the sake of analysis of the differential controllability function, we introduce the following closed-loop system and its prolongation:
\begin{align}\label{eq:dsys_cl}
\left\{\begin{array}{r@{}lr@{}l}
\dot x &{} = f(x) + g (x) k (x), & x(0) &{}= x_0 \\[1mm]
\dot {\delta x} &{} \displaystyle
= \frac{\partial ( f(x) + g(x) k(x))}{\partial x} \delta x, & \delta x(0) &{}= \delta x_0,
\end{array}\right.
\end{align}
where $k: \bR^n \to \bR^m$ is of class $C^2$.
Similar to before, we let $\varphi_k(t, x_0)$ denote the solution to $\dot x = f(x) + g(x) k(x)$, at time $t$, with $x(0)=x_0$.
Then, $(\partial \varphi_k(t, x_0)/ \partial x_0) \delta x_0$ is the solution to the corresponding variational dynamics for $\delta x(0) = \delta x_0$.




\section{Relations between Differential and Incremental Energy Functions}\label{sec:energy}
In this section, we investigate relations between the differential and incremental controllability/observability functions by assuming that each controllability function admits a state feedback solution. In particular, we show that the differential controllability/observability function provides an upper bound on the corresponding incremental function. Conversely, an upper bound on the differential controllability function can be estimated from the incremental controllability function. Moreover, we have a stronger result for observability, in which case the differential observability function can be constructed from the incremental observability function.


\subsection{Incremental Controllability and Observability Functions}
The incremental energy functions are defined as energy functions of the auxiliary system,
\begin{align}\label{eq:isys}
\left\{\begin{array}{r@{}lr@{}l}
\dot x &{} = f(x) + g (x) u, & x(0) &{}= x_0, \\
\dot x' &{} = f(x') + g (x') u', & x'(0) &{}= x'_0, \\
y &{} = h(x) \\
y' &{} = h(x'),
\end{array}\right.
\end{align}
comprising two copies of \eqref{eq:sys}.
Accordingly, the \emph{incremental controllability function} is defined by
\begin{align}\label{eq:icon_func}
E_{i\cC}  (x_0, u, x'_0) := 
\inf_{\substack{u' - u \in L_2^m (-\infty, 0]  \\ x'(-\infty) = x(-\infty)}} \frac{1}{2} \| u' - u \|_{L_2^m(-\infty, 0]}^2
\end{align}
for $x_0 \in \bR^n$, $u: (-\infty, 0] \to \bR^m$, and $x'_0 \in \bR^n$.
Also, the \emph{incremental observability function} is defined by
\begin{align}\label{eq:iob_func}
E_{i\cO} (x_0, x'_0) := \frac{1}{2} \| y' - y \|_{L_2^p[0,\infty)}^2 \; \\
(u(t), u'(t)) \equiv (0, 0)
\nonumber
\end{align}
for $x_0 \in \bR^n$ and $x'_0 \in \bR^n$.

To establish relationships between differential and incremental functions, we employ a modified incremental controllability and observability function. The incremental controllability function is different from that in~\cite[Definition 6]{BWS:14} because we evaluate $u'-u$ instead of $u'+u$ and deal with a $u$-dependent function. The incremental observability function is the special case of~\cite[Definition 5]{BWS:14} where $(u(t), u'(t)) \equiv (0, 0)$.


\subsection{Relations between Controllability Functions}\label{sec:cf}
To connect differential and incremental energy functions, define $\gamma (s) := x_0 + s(x'_0 - x_0)$ and $\bar \gamma (s) := x_0 + s \delta x_0$ for $s \in [0, 1]$. First, we estimate an upper bound on the incremental controllability function on the basis of the differential controllability function.

\begin{secthm}\label{thm:di_con}
For a nonlinear system~\eqref{eq:sys}, suppose that there exist an open convex subset $W \subset \bR^n$ and $k: W \to \bR^m$ of class $C^2$ such that
\begin{enumerate}
\item $W$ is backward complete with respect to $\dot x = f(x) + g(x) k(x)$;
\item for~\eqref{eq:dsys_cl}, there exist $c, \lambda >0$ such that
\begin{align*}
\max\!\left\{ \left|\frac{\partial k(x(t))}{\partial x}\delta x(t)\right|, | \delta x(t) | \right\} \le c e^{ \lambda t} |\delta x_0|
\end{align*}
for all $t \in (-\infty, 0]$ and $(x_0, \delta x_0) \in W \times \bR^n$;
\item the differential controllability function~\eqref{eq:dcon_func} satisfies
\begin{align*}
&E_{d\cC}\bigl(x_0, k(\varphi_k (\cdot , x_0)), \delta x_0\bigr) \\
&\qquad= 
\frac{1}{2} \int_{-\infty}^0 \left| \frac{\partial k(x(t))}{\partial x} \delta x(t) \right|^2 dt
\end{align*}
for any $(x_0, \delta x_0) \in W \times \bR^n$.
\end{enumerate}
Then, the incremental controllability function~\eqref{eq:icon_func} is upper bounded as
\begin{align}\label{eq:di_con}
&\int_0^1 E_{d\cC} \!\left(\gamma (s), k(\varphi_k (\cdot , \gamma (s))), \frac{d \gamma (s)}{ds} \right) ds \nonumber\\
&\qquad\ge E_{i\cC} (x_0, k(\varphi_k (\cdot , x_0)), x'_0)
\end{align}
for any $(x_0, x'_0) \in W \times W$.
\end{secthm}

\begin{proof}
The proof is in Appendix~\ref{app:di_con}.
\end{proof}

Conversely, we can estimate an upper bound on the differential controllability function on the basis of the incremental controllability function as follows. Later, we derive an additional condition for obtaining the equality; see Remark~\ref{rem:id_con} below.

\begin{secthm}\label{thm:id_con}
For a nonlinear system~\eqref{eq:sys}, suppose that there exist a domain (i.e., open and connected) $W \subset \bR^n$ and $k: W \to \bR^m$ of class $C^2$ such that
\begin{enumerate}
\item item 1) in Theorem~\ref{thm:di_con} holds;
\item for \eqref{eq:isys} with $(u, u') = (k(x), k(x'))$, there exist $c, \lambda >0$ such that
\begin{align*}
&\max\bigl\{| k(x'(t)) - k(x(t)) |, | x'(t) - x(t) | \bigr\} \\
&\qquad\le c e^{ \lambda t} | x'_0 - x_0|
\end{align*}
for all $t \in (-\infty, 0]$ and $(x_0, x'_0) \in W \times W$;
\item the incremental controllability function~\eqref{eq:icon_func} satisfies 
\begin{align*}
&E_{i\cC}(x_0, k(\varphi_k (\cdot, x_0)), x'_0)\\
&\qquad= 
\frac{1}{2} \int_{-\infty}^0 | k(x'(t)) - k(x(t)) |^2 dt
\end{align*}
for all $(x_0, x'_0) \in W \times W$.
\end{enumerate}
Then, the differential controllability function~\eqref{eq:icon_func} is upper bounded as
\begin{align}\label{eq:id_con}
&\lim_{s \to 0^+} \frac{E_{i\cC}  (\bar \gamma (0), k(\varphi_k (\cdot , \bar \gamma (0))), \bar \gamma (s))}{s^2} \nonumber\\
&\qquad\ge E_{d\cC} (x_0, k(\varphi_k (\cdot , x_0)), \delta x_0)
\end{align}
for each $(x_0, \delta x_0) \in W \times \bR^n$.
\end{secthm}

\begin{proof}
The proof is in Appendix~\ref{app:id_con}.
\end{proof}


\subsection{Relations between Observability Functions}
In this subsection, we establish similar relations for the observability functions. First, we estimate an upper bound on the incremental observability function on the basis of the differential observability function as follows.

\begin{secthm}\label{thm:di_ob}
For a nonlinear system~\eqref{eq:sys}, suppose that
\begin{enumerate}
\item an open convex subset  $W \subset \bR^n$ is forward complete with respect to $\dot x = f(x)$;
\item for~\eqref{eq:dsys} with $(u(t), \delta u (t)) \equiv (0, 0)$, there exist $c, \lambda >0$ such that
$| \delta y(t) | \le c e^{- \lambda t} |\delta x_0|$ for all $t \in [0, \infty)$ and $(x_0, \delta x_0) \in W \times \bR^n$.
\end{enumerate}
Then, the incremental observability function~\eqref{eq:iob_func} is upper bounded as
\begin{align*}
\int_0^1 E_{d\cO} \!\left( \gamma (s), \frac{d\gamma (s)}{ds} \right) ds \ge E_{i\cO} (x_0, x'_0)
\end{align*}
for each $(x_0, x'_0) \in W \times W$. 
\end{secthm}

\begin{proof}
The proof is in Appendix~\ref{app:di_ob}.
\end{proof}

For the converse, we have a stronger result. In fact, we can construct the differential observability function from the incremental observability function.

\begin{secthm}\label{thm:id_ob}
For a nonlinear system~\eqref{eq:sys}, suppose that
\begin{enumerate}
\item a domain $W \subset \bR^n$ is forward complete with respect to $\dot x = f(x)$;
\item for~\eqref{eq:isys} with $(u(t), u'(t)) \equiv (0, 0)$, there exist $c, \lambda >0$ such that
$| y'(t) - y(t) | \le c e^{- \lambda t} |x'_0 - x_0|$ for all $t \in [0, \infty)$ and $(x_0, x'_0) \in W \times W$.
\end{enumerate}
Then, the differential observability function~\eqref{eq:dob_func} is obtained by
\begin{align*}
\lim_{s \to 0^+} \frac{E_{i\cO} ( \bar \gamma (s), \bar \gamma (0) )}{s^2} = E_{d\cO} (x_0, \delta x_0)
\end{align*}
for each $(x_0, \delta x_0) \in W \times \bR^n$.
\end{secthm}

\begin{proof}
The proof is in Appendix~\ref{app:id_ob}.
\end{proof}




\section{Positive Definiteness of Differential Energy Functions}\label{sec:pd}
In this section, we investigate the positive definiteness of the differential controllability and observability functions from the viewpoint of local accessiblity and observability, respectively. First, we study the differential observability function. Then, we apply the obtained result to analysis of the differential controllability function.

\subsection{Differential Observability Functions}
We have the following characterization for the positive definiteness of the differential observability function, which is a natural extension of the well known result on the observability Gramian of a linear system.

\begin{secthm}\label{thm:ob}
Suppose that item 1) in Theorem~\ref{thm:id_ob} holds, and there exists a (point-wise) symmetric $Q: W \to \bR^{n \times n}$ of class $C^1$ satisfying the differential Lyapunov equation
\begin{align}\label{eq:dLya_ob}
&\sum_{i=1}^n \frac{\partial Q(x)}{\partial x_i} f_i(x)
+ Q(x) \frac{\partial f(x)}{\partial x}  + \frac{\partial^\top f(x)}{\partial x} Q(x) \nonumber\\
&\qquad= - \frac{\partial^\top h(x) }{\partial x} \frac{\partial h(x) }{\partial x}
\end{align}
for all $x \in W$.
Consider the following three properties for~\eqref{eq:dsys} with $(u(t), \delta u(t)) \equiv (0, 0)$:
\begin{enumerate}
\item $\lim_{t \to \infty} \delta x(t) = 0$ for any $(x_0, \delta x_0) \in W \times \bR^n$;
\item zero-state detectability~\cite[Definition 3.2.15]{Schaft:17}, i.e., $\delta y(t) \equiv 0$ implies $\delta x_0 = 0$ for any $x_0 \in W$;
\item $Q(x)$ is the unique symmetric solution to~\eqref{eq:dLya_ob} and positive definite on $W$, and the differential observability function~\eqref{eq:dob_func} satisfies
\begin{align*}
E_{d\cO} (x_0, \delta x_0) = \frac{1}{2} |\delta x_0|_{Q(x_0)}^2
\end{align*}
for all $(x_0, \delta x_0) \in W \times \bR^n$.
\end{enumerate}
Then, we have the following implications:
\begin{itemize}
\item items 1) and 2) imply item 3); 
\item items 2) and 3) imply item 1);
\item items 3) and 1) imply item 2). 
\end{itemize}
\end{secthm}

\begin{proof}
The proof is in Appendix~\ref{app:ob}.
\end{proof}

\begin{secrem}
According to the proof of Theorem~\ref{thm:ob}, in particular \eqref{pf1:dob_func} below, we have
\begin{align}\label{eq:dobG}
Q(x) = \int_0^\infty \left(\frac{\partial h(\varphi)}{\partial \varphi} \frac{\partial \varphi}{\partial x}\right)^\top \frac{\partial h(\varphi)}{\partial \varphi} \frac{\partial \varphi }{\partial x} dt
\end{align}
for all $x \in W$, where the argument of $\varphi$ is $(t,x,0)$. Thus, $Q(x)$ is nothing but the \emph{differential observability Gramian} \cite[Definition 3.2]{KS:21}.
\red
\end{secrem}

\begin{secrem}\label{rem:IAS}
If item 1) in Theorem~\ref{thm:di_ob} and items 2) and~3) of Theorem~\ref{thm:ob} hold, then the LaSalle invariance principle for contraction analysis~\cite[Theorem 2]{FS:14} concludes that $\dot x = f(x)$ is incrementally asymptotically stable~\cite[Definition 1]{FS:14} on $W$.
\red
\end{secrem}

\begin{secrem}\label{rem:lob}
Let $f(x)$ and $h(x)$ be (real) analytic on $W$. Then, item 2) of Theorem~\ref{thm:ob} is equivalent to that the system~\eqref{eq:sys} with $u = 0$ satisfies the observability rank condition on $W$, e.g.,~\cite[Theorem 3.32]{NS:15}. This can be shown by computing the time derivative of $\delta y(t)$ based on the convergence of the Taylor expansion of $\delta y(t)$ with respect to $t \in \bR$. 
\red
\end{secrem}




\subsection{Differential Controllability Functions}
In this subsection, we study positive definiteness of the differential controllability function. In \cite[Theorem 2.5]{KS:17}, this function is implicitly assumed to be independent of $u$. In this paper, we handle the case where $u = k(x)$ as in Section~\ref{sec:cf}. In this case, the differential controllability function is obtained by solving a differential Riccati equation, stated below.

\begin{secthm}\label{thm:con}
For a nonlinear system~\eqref{eq:sys}, suppose that there exist a domain $W \subset \bR^n$, a symmetric $R: W \to \bR^{n \times n}$ of class $C^1$, and $k: W \to \bR^m$ of class $C^2$ such that item~1) in Theorem~\ref{thm:di_con} and
\begin{subequations}\label{eq:dRicc_con}
\begin{align}
&\sum_{i=1}^n \frac{\partial R(x)}{\partial x_i} (f(x) + g(x) k(x))_i \nonumber\\
& \; + R(x) \frac{\partial (f(x) + g(x) k(x))}{\partial x} + \frac{\partial^\top (f(x) + g(x) k(x))}{\partial x} R(x) \nonumber\\
&= \frac{\partial^\top k(x)}{\partial x} \frac{\partial k(x)}{\partial x},
\label{eq1:dRicc_con}\\
&\frac{\partial k(x)}{\partial x} = g^\top (x) R(x)
\label{eq2:dRicc_con}
\end{align}
hold for all $x \in W$.
\end{subequations}
Then, the differential controllability function~\eqref{eq:dcon_func} satisfies 
\begin{align*}
E_{d\cC} \bigl(x_0, k(\varphi_k (\cdot , x_0)), \delta x_0\bigr) = \frac{1}{2} |\delta x_0|_{R(x_0)}^2
\end{align*}
for any $(x_0, \delta x_0) \in W \times \bR^n$ such that~\eqref{eq:dsys_cl} fulfills $\lim_{t \to -\infty} \delta x(t) = 0$. 
\end{secthm}

\begin{proof}
The proof is in Appendix~\ref{app:con}.
\end{proof}

\begin{secrem}\label{rem:id_con}
One notices that if $k(x)$ satisfies the conditions in both Theorems~\ref{thm:id_con} and \ref{thm:con}, then equality holds in~\eqref{eq:id_con}. Moreover, we have
\begin{align*}
R(x) = \int_{-\infty}^0 \left(\frac{\partial k(\varphi_k)}{\partial \varphi_k} \frac{\partial \varphi_k}{\partial x}\right)^\top \frac{\partial k(\varphi_k)}{\partial \varphi_k} \frac{\partial \varphi_k}{\partial x} dt,
\end{align*}
where the argument of $\varphi_k$ is $(t,x)$.
\red
\end{secrem}

Suppose that $R(x)$ is symmetric and positive definite at each $x \in W$. Let $P(x)$ denote the inverse of $R(x)$, i.e., $P(x):=R^{-1}(x)$. Then,~\eqref{eq:dRicc_con} is equivalent to
\begin{subequations}\label{eq:dLya_con}
\begin{align}
&- \sum_{i=1}^n \frac{\partial P(x)}{\partial x_i} (f(x) + g(x) k(x))_i \nonumber\\
&\quad + P(x) \left.\frac{\partial^\top (f(x) + g(x) u)}{\partial x}\right|_{u=k(x)} \nonumber\\
&\quad + \left.\frac{\partial (f(x) + g(x) u)}{\partial x}\right|_{u=k(x)} P(x) 
= -g(x) g^\top (x),
\label{eq1:dLya_con}\\
&\frac{\partial k(x)}{\partial x}P(x) = g^\top (x).
\end{align}
\end{subequations}
One notices that~\eqref{eq1:dLya_con} is nothing but~\eqref{eq:dLya_ob} for the following system:
\begin{align}\label{eq:sys_z}
\left\{\begin{array}{r@{}lr@{}l}
\dot x &{} = -(f(x) + g (x) k(x))\\[1mm]
\dot {\delta p} &{}\displaystyle = \left.\frac{\partial^\top (f(x) + g (x) u)}{\partial x}\right|_{u=k(x)} \delta p\\[2mm]
\delta z &{} = g^\top (x) \delta p, \quad
(x(0), \delta p(0)) = (x_0, \delta p_0).
\end{array}\right.
\end{align}
Thus, applying Theorem~\ref{thm:ob} to analyzing the positive definiteness of $P(x)$ yields the following corollary.

\begin{seccor}\label{cor:con}
For a nonlinear system~\eqref{eq:sys}, suppose that there exist a domain $W \subset \bR^n$, symmetric $P: W \to \bR^{n \times n}$ of class $C^1$, and $k: W \to \bR^m$ of class $C^2$ such that item~1) in Theorem~\ref{thm:di_con} and \eqref{eq:dLya_con} hold. Consider the following three properties for \eqref{eq:sys_z}:
\begin{enumerate}
\item $\lim_{t \to \infty} \delta p (t) = 0$ for any $(x_0, \delta p_0) \in W \times \bR^n$;
\item $\delta z(t) = 0$ for any $t \in [0, \infty)$ implies $\delta p_0 = 0$ for any $x_0 \in W$;
\item $P(x)$ is positive definite on $W$.
\end{enumerate}
Then, we have the following implications:
\begin{itemize}
\item items 1) and 2) imply item 3); 
\item items 2) and 3) imply item 1);
\item items 3) and 1) imply item 2). 
\end{itemize}
\end{seccor}
\begin{proof}
The proof is similar to that of Theorem~\ref{thm:ob}.
\end{proof}

Let $f(x)$, $g(x)$, and $k(x)$ be analytic on $W$. Applying the discussion of Remark~\ref{rem:lob}, item 2) of Corollary~\ref{cor:con} is equivalent to that the system~\eqref{eq:sys_z} satisfies 
\begin{align*}
\left.\frac{d^i \delta z(t)}{dt^i} \right|_{t=0} = 0, \quad i=0,1,\dots
\end{align*}
for all $(x, \delta p_0) \in W \times \bR^n$. This is equivalent to
\begin{align}\label{eq:lac}
{\rm rank} \begin{bmatrix} g \!& {\rm ad}_{f+gu| k} g \!& \cdots \!& {\rm ad}_{f+gu| k}^i g \!& \cdots \; \end{bmatrix}(x) = n
\end{align}
for all $x \in W$, where ${\rm ad}_{f+gu| k}^0 g(x) := g(x)$ and, for $i = 0, 1, \dots$,
\begin{align*}
{\rm ad}_{f+gu| k}^{i+1} g(x) 
&:= \frac{\partial {\rm ad}_{f+gu| k}^i g(x)}{\partial x} (f(x) + g(x) k(x)) \\
&\phantom{:=}\; - \left.\frac{\partial (f(x)\!+\!g(x)u)}{\partial x}\right|_{u=k(x)} {\rm ad}_{f+gu| k}^i g(x).
\end{align*}
This is different from ${\rm ad}_{f+gk}^i g(x)$ that is commonly seen in differential geometric nonlinear control theory, e.g.,~\cite{NS:15}.

Corollary~\ref{cor:con} recovers a well-known result on the controllability Gramian of a linear system. 
Let $f(x) = A x$, $g(x) = B$, and $k(x) = K x$.
First, item 1) requires the exponential stability of $\dot p = A^\top p$, i.e., $\dot x = A x$. Regarding item~2), the condition~\eqref{eq:lac} becomes the controllability rank condition of the pair $(A, B)$. Finally, item~3) is nothing but the positive definiteness of the controllability Gramian.

Corollary~\ref{cor:con} deals with the controlled system~\eqref{eq:sys_z}. When one is interested in  an open nonlinear system~\eqref{eq:sys}, one can utilize the differential Lyapunov equation
\begin{align}\label{eq3:dLya_con}
&- \sum_{i=1}^n \frac{\partial \overline P(x)}{\partial x_i} f_i(x) + \overline P(x) \frac{\partial^\top f(x)}{\partial x} + \frac{\partial f(x)}{\partial x} \overline P(x) \nonumber\\
&\qquad= -g(x) g^\top (x)
\end{align}
and the dynamics
\begin{align}\label{eq:sys_z2}
\left\{\begin{array}{r@{}lr@{}l}
\dot x &{} = -f(x)\\[1mm]
\dot {\delta p} &{}\displaystyle = \frac{\partial^\top f(x)}{\partial x} \delta p\\[2mm]
\delta z &{} = g^\top (x) \delta p, \quad
(x(0), \delta p(0)) = (x_0, \delta p_0).
\end{array}\right.
\end{align}
Then, similar implications as in Corollary~\ref{cor:con} hold. 

Moreover, the counterparts of items 1) and 2) can be rewritten as follows.
From $x(t) = \varphi (t, x, 0)$, one can confirm that
\begin{align*}
\delta p(t) = \left(\frac{\partial \varphi (-t, x_0, 0)}{\partial x_0}\right)^{-\top} \delta p_0
\end{align*}
satisfies~\eqref{eq:sys_z2}.
Applying \cite[Theorem 1.42]{KS:72} concludes that the exponential convergence of $\delta p(t)$ and $\delta x(t)$ in~\eqref{eq:dsys} with $(u(t), \delta u(t)) \equiv (0, 0)$ (both forward in time) are equivalent. Namely, a sufficient condition for item 1) can be derived as a condition for the system~\eqref{eq:dsys}.
The counterpart of~\eqref{eq:lac} is
\begin{align}\label{eq:lac2}
{\rm rank} \begin{bmatrix} g & {\rm ad}_f g & \cdots & {\rm ad}_f^i g & \cdots \; \end{bmatrix}(x) = n,
\end{align}
which is a sufficient condition for the local strong accessibility rank condition~\cite[Proposition 3.20]{NS:15} for the system~\eqref{eq:dsys}. The rank condition~\eqref{eq:lac2} is equivalent to the complete controllability~\cite[Corollary 1]{SM:67} of the variational system
\begin{align*}
\dot {\overline{\delta x}} = \frac{\partial f(\varphi(t, x, 0))}{\partial \varphi} \overline{\delta x} + g(\varphi(t, x, 0)) \overline{\delta u},
\end{align*}
in the linear time-varying sense.

As a final remark, \eqref{eq3:dLya_con} appears in the analysis of the $u$-independent differential controllability function \cite[Theorem 2.5]{KS:17}, where the so-called killing vector field conditions with respect to $g_i(x)$, $i=1,\dots,m$ are additionally required. Therefore, a stronger property than local strong accessibility can be instrumental to understand the positive definiteness of the differential controllability functions, regardless of the dependence on $u$.




\section{Example}\label{sec:ex}
Consider the nonlinear system~\eqref{eq:sys} with
\begin{align}
f(x) &= 
\begin{bmatrix}
-x_1/2 - x_1^2 - x_1^3/3 - x_1 x_2 - x_2\\
-x_2/2
\end{bmatrix}\nonumber\\
g(x) &=
\begin{bmatrix}
1+x_1\\
1
\end{bmatrix},
\quad
h(x) = x_1.
\label{eq:ex}
\end{align}

For controllablility, a solution to~\eqref{eq:dLya_con} is
\begin{align*}
P(x) = I_2 \succ 0, \quad
k(x) = x_1 + \frac{x_1^2}{2} + x_2.
\end{align*}
Thus, item 3) in Corollary~\ref{cor:con} holds.

\begin{figure}[t]
\centering
\includegraphics[width=0.9\columnwidth]{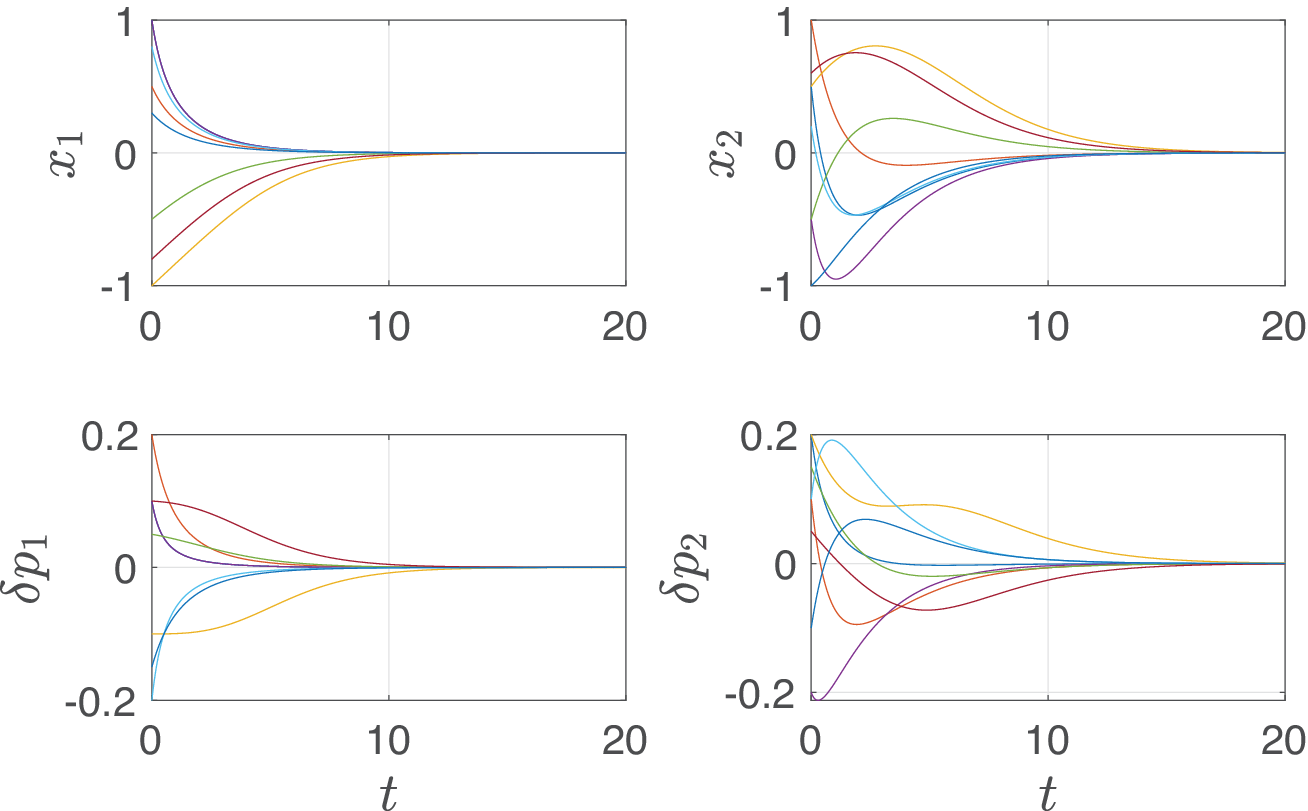}
\caption{Time response of~\eqref{eq:sys_z} with~\eqref{eq:ex}}
\label{fig:sim}
\end{figure}

We check item 2). It follows that
\begin{align*}
&{\rm ad}_{f+gu| k} g(x) 
= \begin{bmatrix}
3/2 + 2 x_1 + x_1^2 + x_1^3/3 \\ 1/2   
    \end{bmatrix}\\
&{\rm ad}_{f+gu| k}^2 g(x) \\
&= \begin{bmatrix}
5/4 + 4 x_1 + 21 x_1^2/4 + 4 x_1^3 + 5 x_1^4/3 + x_1^5/3 \\ 1/4   
    \end{bmatrix}
\end{align*}
One can confirm that~\eqref{eq:lac}, i.e., item 2) in Corollary~\ref{cor:con}, holds for all $x \in \bR^2$. According to Corollary~\ref{cor:con}, item 1) holds, which is validated by simulation in Fig.~\ref{fig:sim}.

\begin{figure}[t]
\centering
\includegraphics[width=0.9\columnwidth]{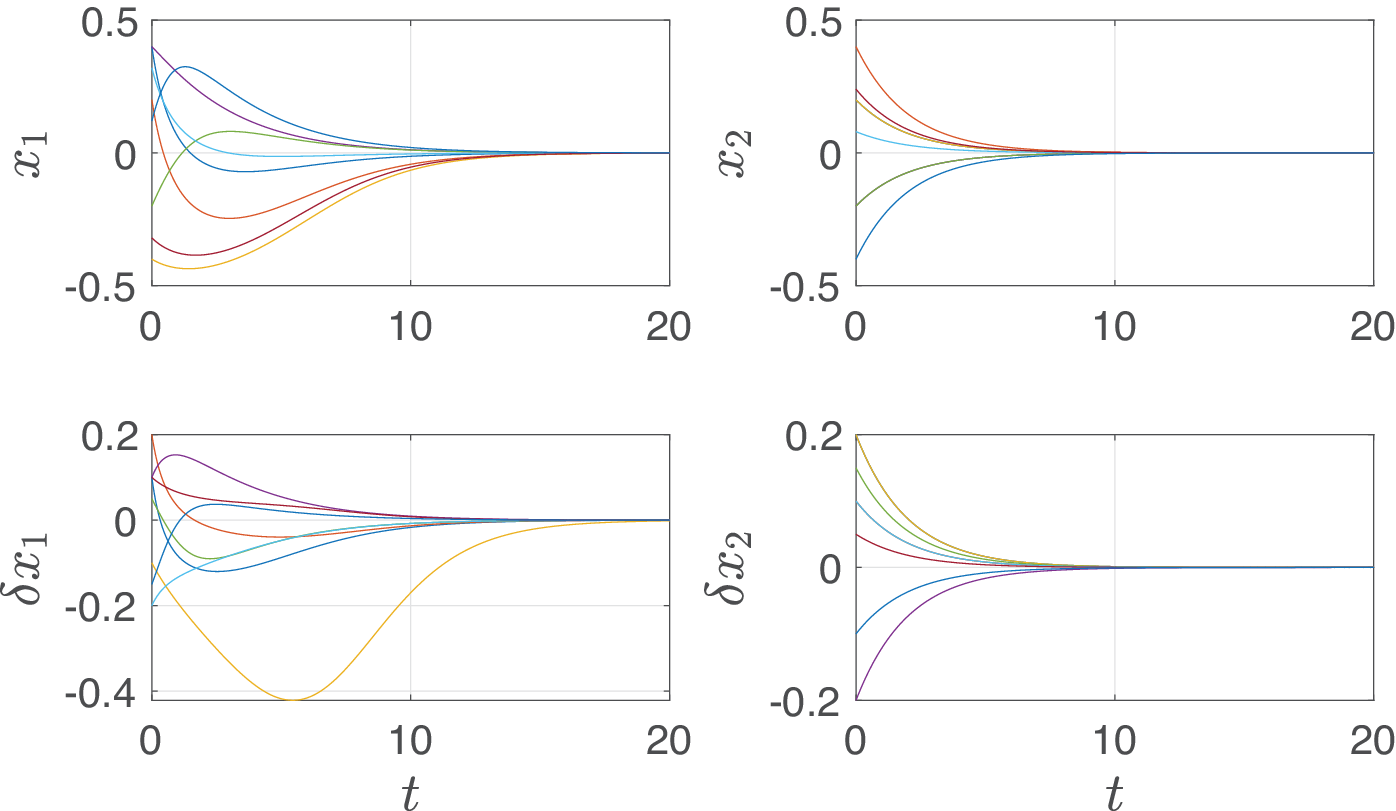}
\caption{Time response of~\eqref{eq:dsys} with~\eqref{eq:ex} when $(u(t), \delta u(t))\equiv (0,0)$}
\label{fig:sim2}
\end{figure}

\begin{figure}[t]
\centering
\includegraphics[width=0.6\columnwidth]{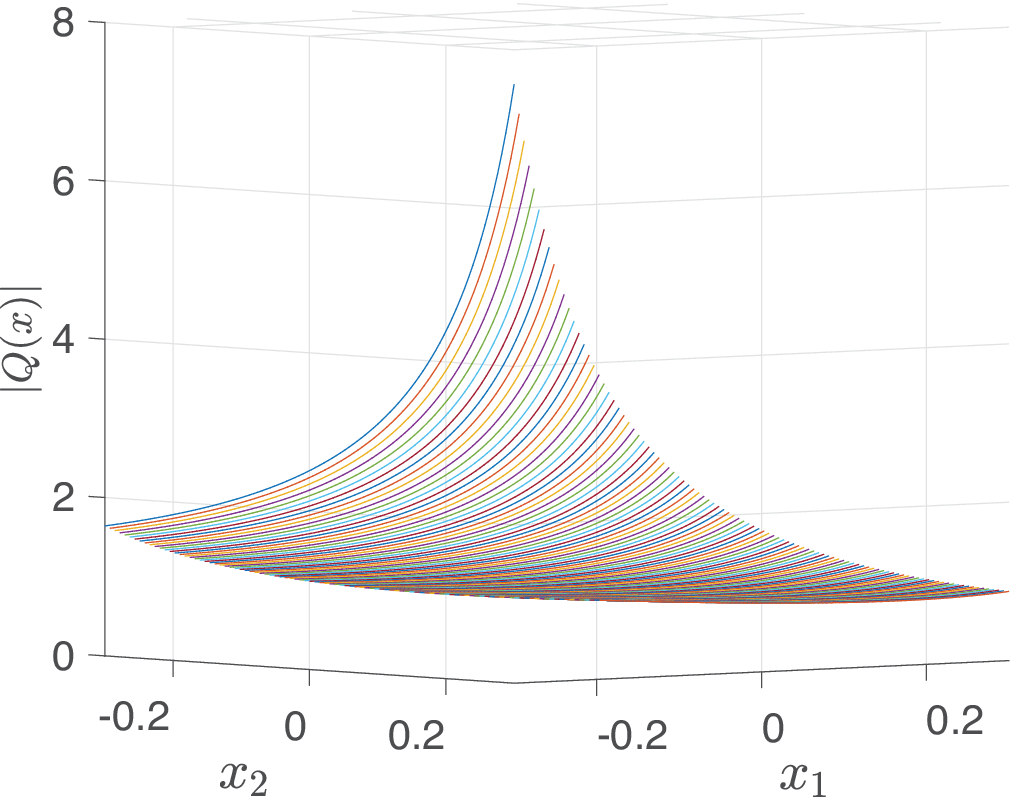}
\caption{$|Q(x)|$}
\label{fig:Q}
\end{figure}

Next, we consider observability. Since the Jacobian matrix at the origin, i.e., $\partial f(0)/\partial x$, is Hurwitz for~\eqref{eq:ex}, there exists a forward complete set $W \subset \bR^2$ containing the origin such that item 1) of Theorem~\ref{thm:ob} holds as in Fig.~\ref{fig:sim2}. Since the convergence of both $x(t)$ and $\delta x(t)$ are exponential around the origin, the differential observability Gramian $Q(x)$ in~\eqref{eq:dobG} exists at least around $x=0$.

Based on Remark~\ref{rem:lob}, we verify item 2). It follows that
\begin{align*}
&{\rm rank}
\begin{bmatrix}
\partial h/\partial x_1 & \partial h/\partial x_2\\
\partial L_fh/\partial x_1 & \partial L_fh/\partial x_2
\end{bmatrix}(x)\\
&={\rm rank}
\begin{bmatrix}
1 & 0\\
-1/2 - 2x_1 - 3 x_1^2 - x_2 & -1 - x_1
\end{bmatrix},
\end{align*}
where $L_f h(x) := (\partial h(x)/\partial x)f(x)$. Thus, the observability rank condition holds for $x_1 \neq -1$. According to Theorem~\ref{thm:ob}, the differential observability Gramian $Q(x)$ is positive definite at least around $x=0$. To confirm this, Fig.~\ref{fig:Q} shows the determinant of $Q(x)$, which is positive for all $x \in [-0.3, 0.3]^2$.




\section{Conclusion}\label{sec:con}
In this paper, we have deepened the understanding of differential controllability and observability functions from two distinct perspectives. First, we have demonstrated that an upper bound on each differential energy function can be derived from the corresponding incremental function, and vice versa. In the case of observability, we have specifically showed that the differential observability function can be constructed from the incremental observability function. Second, we have studied the positive definitenss of both differential energy functions. For the differential observability function, we have established natural extensions of the well known relations, in the linear case, among exponential stability, observability, and the observability Gramian, by utilizing the convergence of the variational dynamics and local observability. Finally, we have applied this result to analysis of the differential controllability Gramian. In contrast to observability, we have found that a property stronger than local strong accessibility is generally required to ensure the positive definiteness of the differential controllability function, which remains a space of further investigation. 




\appendix
\subsection{Proofs of Theorems in Section~\ref{sec:energy}}
\subsubsection{Proof of Theorem~\ref{thm:di_con}}\label{app:di_con}
According to the proof of~\cite[Theorem 5.4]{KB:24},
\begin{align}\label{pf3:di_con}
(x(t), \delta x(t)) 
= 
\left(\varphi_k (t, \gamma (s)), \frac{d \varphi_k (t, \gamma (s))}{d s} \right)
\end{align}
satisfies \eqref{eq:dsys_cl} for any $s \in [0, 1]$ and $(x_0, x'_0) \in W \times W$. Thus, it follows from items 1) and 3) that
\begin{align*}
&E_{d\cC} \left(\gamma (s), k(\varphi_k (\cdot , \gamma (s))), \frac{d \gamma (s)}{ds} \right) \\
&\qquad= \frac{1}{2} \int_{-\infty}^0  \left| \frac{d k(\varphi_k (t, \gamma (s)))}{d s} \right|^2 dt.
\end{align*}
Integration with respect to $s$ over the interval $[0,1]$ yields
\begin{align*}
&\int_0^1 E_{d\cC} \left( \gamma (s), k(\varphi_k (\cdot , \gamma (s))), \frac{d \gamma (s)}{ds} \right) ds \\
&\qquad= \frac{1}{2} \int_0^1 \int_{-\infty}^0  \left| \frac{d k(\varphi_k (t, \gamma (s)))}{d s} \right|^2 dt ds\\
&\qquad= \frac{1}{2} \int_{-\infty}^0  \int_0^1  \left| \frac{d k(\varphi_k (t, \gamma (s)))}{d s} \right|^2 ds dt\\
&\qquad\ge \frac{1}{2} \int_{-\infty}^0  \left| \int_0^1  \frac{d k(\varphi_k (t, \gamma (s)))}{d s} ds \right|^2  dt\\
&\qquad= \frac{1}{2} \int_{-\infty}^0  |k(\varphi_k (t, x'_0)) - k(\varphi_k (t, x_0))|^2  dt,
\end{align*}
where the second equality follows from item 2).

To show~\eqref{eq:di_con}, it remains to prove that $k(\varphi_k (t, x'_0))$ achieves $x'(-\infty) = x(-\infty)$.
To do so, note that items 1) and 2) imply
\begin{align*}
&\limsup_{t \to -\infty} | \varphi_k (t, x'_0) - \varphi_k (t, x_0) | \\
&\qquad= \limsup_{t \to -\infty} \left| \int_0^1 \frac{d \varphi_k (t, \gamma (s))}{d s} ds \right|\\
&\qquad\le \limsup_{t \to -\infty} \int_0^1 \left| \frac{d \varphi_k (t, \gamma (s))}{d s} \right| ds\\
&\qquad\le \limsup_{t \to -\infty} c e^{\lambda t}  \int_0^1 \left| \frac{d\gamma (s)}{ds}\right|  ds  = 0,
\end{align*}
which completes the proof.
\QED



\subsubsection{Proof of Theorem~\ref{thm:id_con}}\label{app:id_con}
Substituting $(x_0, u) = (\bar \gamma (0), k(\varphi_k (\cdot,\bar \gamma (0))))$ and $(x'_0, u') = (\bar \gamma (s), k(\varphi_k (\cdot, \bar \gamma (s))))$ into the equality of item~3) and using item~1) leads to
\begin{align*}
&E_{i\cC}  (\bar \gamma (0), k(\varphi_k (\cdot,\bar \gamma (0))), \bar \gamma (s) ) \\
&\qquad= \frac{1}{2} \int_{-\infty}^0 |k(\varphi_k (t, \bar \gamma (s))) - k(\varphi_k (t,\bar \gamma (0)))|^2 dt,
\end{align*}
and, consequently,
\begin{align}\label{pf1:id_con}
&\limsup_{s \to 0^+} \frac{E_{i\cC}  (\bar \gamma (0), k(\varphi_k (\cdot,\bar \gamma (0))), \bar \gamma (s) )}{s^2} \nonumber\\
&= \frac{1}{2} \limsup_{s \to 0^+} \int_{-\infty}^0 \left| \frac{k(\varphi_k (t,\bar \gamma (s))) - k(\varphi_k (t,\bar \gamma (0)))}{s} \right|^2 dt \nonumber\\
&=  \frac{1}{2} \int_{-\infty}^0 \left| \limsup_{s \to 0^+} \frac{k(\varphi_k (t,\bar \gamma (s))) - k(\varphi_k (t,\bar \gamma (0)))}{s} \right|^2 dt \nonumber\\
&=  \frac{1}{2} \int_{-\infty}^0 \left| \frac{d k(\varphi_k (t,\bar \gamma (0)))}{ds}  \right|^2 dt \nonumber\\
&=  \frac{1}{2} \int_{-\infty}^0 \left| \frac{\partial k(\varphi_k (t,x_0))}{\partial \varphi_k} \frac{\partial \varphi_k(t, x_0)}{\partial x_0} \delta x_0 \right|^2 dt,
\end{align}
where the second equality follows from item 2). Also, the third equality implies that the limit superior can be replaced with the ordinary limit.

Finally, we confirm that $\delta u(t) = \partial \varphi_k (t, x_0)/\partial x_0$ achieves $\delta x(-\infty) = 0$. Namely, it follows from items 1) and~2) that
\begin{align*}
&\limsup_{t \to -\infty} \left| \frac{\partial \varphi_k(t, x_0)}{\partial x_0} \delta x_0 \right|\\
&\qquad=\limsup_{t \to -\infty} \left| \frac{d \varphi_k (t, \bar \gamma (0))}{ds} \right| \\
&\qquad=\limsup_{t \to -\infty} \left|  \lim_{s \to 0^+} \frac{\varphi_k (t, \bar \gamma (s)) - \varphi_k (t,\bar \gamma (0))}{s}\right| \\
&\qquad=\limsup_{t \to -\infty}\lim_{s \to 0^+}  \left|  \frac{\varphi_k (t, \bar \gamma (s)) - \varphi_k (t,\bar \gamma (0))}{s}\right| \\
&\qquad\le \limsup_{t \to -\infty} c e^{\lambda t} | \delta x_0 | = 0.
\end{align*}
This completes the proof.
\QED

Although \eqref{pf1:id_con} does not contain any inequality, this is not enough to guarantee that $\delta u = (\partial k(x)/\partial x)\delta x$ achieves the infimum~\eqref{eq:dcon_func} for $u = k(x)$. This is the reason why we do not have equality in~\eqref{eq:id_con}. 


\subsubsection{Proof of Theorem~\ref{thm:di_ob}}\label{app:di_ob}
According to the proof of~\cite[Theorem 5.4]{KB:24},
\begin{align*}
(x(t), y(t)) &= (\varphi (t, \gamma (s), 0), h(\varphi (t, \gamma (s), 0)) ) \\[1mm]
(\delta x(t), \delta y(t)) &= \left(\frac{d\varphi (t, \gamma (s), 0)}{ds} , \frac{dh(\varphi (t, \gamma (s), 0))}{ds} \right)
\end{align*}
satisfy~\eqref{eq:dsys} with $(u(t), \delta u(t)) \equiv (0,0)$ for any $s \in [0, 1]$ and $(x_0, x'_0) \in W \times W$.
Thus, items 1) and 2) lead to
\begin{align*}
&\int_0^1 E_{d\cO} \left( \gamma (s), \frac{d\gamma (s)}{ds}\right) ds \\
&\qquad= \frac{1}{2} \int_0^1  \int_0^\infty \left| \frac{d h(\varphi(t,\gamma(s),0))}{ds} \right|^2 dt ds \\
&\qquad= \frac{1}{2} \int_0^\infty \int_0^1  \left| \frac{d h(\varphi(t,\gamma(s),0))}{ds} \right|^2 ds dt\\
&\qquad\ge \frac{1}{2} \int_0^\infty  \left| \int_0^1 \frac{d h(\varphi(t,\gamma(s),0))}{ds}  ds\right|^2  dt \\
&\qquad= \frac{1}{2} \int_0^\infty |h(\varphi(t,x'_0,0)) - h(\varphi(t,x_0,0)) |^2 dt \\
&\qquad= E_{i\cO} (x_0, x'_0),
\end{align*}
where the second equality follows from item 2). 
\QED


\subsubsection{Proof of Theorem~\ref{thm:id_ob}}\label{app:id_ob}
Substituting $(x_0, u) = (\bar \gamma (0), 0)$ and $(x'_0, u') = (\bar \gamma (s), 0)$ into $E_{i\cO}(x_0, x'_0)$ yields
\begin{align*}
&E_{i\cO} (\bar \gamma (s), \bar \gamma (0))\\
&\qquad= \frac{1}{2}  \int_0^\infty | h(\varphi(t, \bar \gamma (s), 0)) - h(\varphi(t, \bar\gamma (0), 0)) |^2 dt,
\end{align*}
and consequently,
\begin{align*}
&\limsup_{s \to 0^+} \frac{E_{i\cO} (\gamma (s), \gamma (0))}{s^2}\\
&= \frac{1}{2} \limsup_{s \to 0^+} \int_0^\infty \left| \frac{h(\varphi(t, \gamma (s), 0)) - h(\varphi(t, \gamma (0), 0))}{s} \right|^2 dt \\
&= \frac{1}{2} \int_0^\infty \left| \limsup_{s \to 0^+} \frac{h(\varphi(t, \gamma (s), 0)) - h(\varphi(t, \gamma (0), 0))}{s} \right| dt \\
&= \frac{1}{2} \int_0^\infty \left| \frac{d h(\varphi(t, \gamma (0), 0))}{ds} \right| dt \\
&= \frac{1}{2} \int_0^\infty \left| \frac{\partial h(\varphi(t, x_0, 0))}{\partial \varphi} \frac{\partial \varphi(t, x_0, 0)}{\partial x_0} \delta x_0 \right| dt\\
&= E_{d\cO} (x_0, \delta x_0),
\end{align*}
where the second equality follows from item 2). Also, the third equality implies that the limit superior can be replaced with the ordinary limit.
\QED



\subsection{Proofs of Theorems in Section~\ref{sec:pd}}
\subsubsection{Proof of Theorem~\ref{thm:ob}}\label{app:ob}
(items 1) and 2) $\implies$ item 3))
According to the proof of \cite[Theorem 2.7]{KS:17} and linearity of the variational dynamics, item 1) implies
\begin{align}\label{pf1:dob_func}
\|\delta y \|_{L_2^p [0, \infty)}^2 = |\delta x_0|_{Q(x_0)}^2
\end{align}
for all $(x_0, \delta x_0) \in W \times \bR^n$.
Given $(x_0, \delta x_0) \in W \times \bR^n$, $\delta y(t)$ is uniquely determined. Since all symmetric solutions to~\eqref{eq:dLya_ob} satisfy~\eqref{pf1:dob_func}, $Q(x)$ is the unique symmetric solution. Finally, from item 2) and the continuity of $\delta y(t)$ with respect to $t \in \bR$, if $\delta x_0 \neq 0$, then $\delta y(t) \neq 0$ in some time-interval, and thus $Q(x_0) \succ 0$.

(items 2) and 3) $\implies$ item 1))
Let $L_\cO (x, \delta x):= |\delta x|_{Q(x)}^2$. From \cite[Theorem 2]{FS:14} and \eqref{eq:dLya_ob}, all trajectories of~\eqref{eq:dsys} starting from $(x(0), \delta x(0)) \in W \times \bR^n$ with $(u(t), \delta u(t)) \equiv (0,0)$ converge to the largest invariant set $M$ contained in $\{(x, \delta x) \in W \times \bR^n : \delta y = 0\}$. By the invariance principle, e.g., \cite[Theorem 4.4]{Khalil:02}, item~2) implies item~1).

(items 3) and 1) $\implies$ item 2))
Item 1) yields~\eqref{pf1:dob_func}. From item 3), $\delta x_0 \neq 0$ implies $\|\delta y \|_{L_2^p [0, \infty)}^2 \neq 0$. By the continuity of $\delta y(t)$ with respect to $t \in \bR$, we have item 2).
\QED



\subsubsection{Proof of Theorem~\ref{thm:con}}\label{app:con}
The variational system~\eqref{eq:dsys} along the trajectory $(x(t), u(t))=(\varphi_k (t,x_0), k(\varphi_k (t,x_0)))$ is
\begin{align}\label{pf1:dcon_func}
&\dot {\delta x} 
= F(\varphi_k (t,x_0), k(\varphi_k (t,x_0))) \delta x 
+ g(\varphi_k (t,x_0)) \delta u,
\end{align}
where
\begin{align*}
F(x, u) : = 
\frac{\partial f(x)}{\partial x}
+ \sum_{j=1}^m \frac{\partial g_j (x)}{\partial x_j} u_j.
\end{align*}
By the backward completeness of $W$, $\delta x(t)$ exists for all $t \in (-\infty, 0]$, $(x_0, \delta x_0) \in W \times \bR^n$, and $\delta u \in L_2^m(-\infty, 0]$.

Define $\bar f(x) := f(x) + g(x) k(x)$.
From~\eqref{eq:dRicc_con}, the time derivative of $|\delta x(t)|_{R(\varphi_k (t,x_0))}^2$ along trajectories of~\eqref{pf1:dcon_func} satisfies
\begin{align*}
&\frac{d}{dt} |\delta x(t)|_{R (\varphi_k (t,x_0))}^2\\
&= \delta x^\top (t) \sum_{i=1}^n \frac{\partial R(\varphi_k (t,x_0))}{\partial \varphi_{k,i}} \bar f_i(\varphi_k (t,x_0)) \delta x(t) \\
&\quad 
+ 2 \delta x^\top (t) R(\varphi_k (t,x_0)) F(\varphi_k (t,x_0), k(\varphi_k (t,x_0)))  \delta x (t)\\
&\quad 
+ 2 \delta x^\top (t) R(\varphi_k (t,x_0)) g(\varphi_k (t,x_0)) \delta u (t) \\
&= |\delta u(t)|^2 - \left| \delta u(t) - \frac{\partial k(\varphi_k (t,x_0))}{\partial \varphi_k} \delta x (t) \right|^2.
\end{align*}
Integration of this result over the time interval $(-\infty, 0]$ yields
\begin{align*}
&|\delta x_0|_{R (x_0)}^2 - \lim_{t \to -\infty} |\delta x(t)|_{R (x(t))}^2 \\
&= \| \delta u \|_{L_2^m(-\infty, 0]}^2 - \int_{-\infty}^0 \left| \delta u(t) - \frac{\partial k(\varphi_k (t,x_0))}{\partial \varphi_k} \delta x (t) \right|^2 dt \\
&\le \| \delta u \|_{L_2^m(-\infty, 0]}^2 .
\end{align*}
This inequality becomes the equality if $\delta u(t) = (\partial k(\varphi_k (t,x_0))/\partial \varphi_k) \delta x(t)$. The system~\eqref{pf1:dcon_func} with this $\delta u$ is nothing but~\eqref{eq:dsys_cl}. Thus, $\lim_{t \to -\infty} \delta x(t) = 0$ for the considered $(x_0, \delta x_0) \in W \times \bR^n$. Hence, it follows that
\begin{align*}
|\delta x_0|_{R (x_0)}^2 
&= \int_{-\infty}^0 \left| \frac{\partial k(\varphi_k (t,x_0))}{\partial \varphi_k} \delta x(t) \right|^2 dt \\
&\le \| \delta u \|_{L_2^m(-\infty, 0]}^2.
\end{align*}
Note that this holds for any $\delta u \in L_2^m(-\infty, 0]$, implying
\begin{align*}
|\delta x_0|_{R (x_0)}^2 
&= \int_{-\infty}^0 \left| \frac{\partial k(\varphi_k (t,x_0))}{\partial \varphi_k} \delta x(t) \right|^2 dt \\
&= \inf_{\substack{\delta u \in L_2(-\infty, 0]  \\ \delta x(-\infty)=0}} \| \delta u \|_{L_2^m(-\infty, 0]}^2.
\end{align*}
This completes the proof.

\bibliographystyle{IEEEtran}
\bibliography{ref}

\end{document}